\documentclass[%
superscriptaddress,%
twocolumn,%
prb,%
aps,%
showpacs,%
preprintnumbers,%
amsmath,%
amssymb,%
twoside,%
floatfix]{revtex4-1}

\usepackage{graphicx}		% Include figure files
\usepackage{dcolumn}		% Align table columns on decimal point
\usepackage{color}	    	% text color possible
\usepackage[colorlinks=true,
            citecolor=blue, 
            linkcolor=blue,
            urlcolor=blue,
            pdfborder={0 0 0}]{hyperref}

\begin{document}
%
% reference to a figure
\newcommand{\myref}[1]{Fig.\,\ref{#1}}

%
% modulation vectors
\newcommand{\qvec}{$\mathbf{q}$-vector}

\newcommand{\cqn}{$(3/13, 1/13)$ }
\newcommand{\cqs}{$\mathbf{q}_\mathrm{C}$}

\newcommand{\ncqn}{$(0.248, 0.068, 1/3)$ }
\newcommand{\ncqs}{$\mathbf{q}_\mathrm{NC}$}

\newcommand{\icqn}{$(0.28, 0, 1/3)$ }
\newcommand{\icqs}{$\mathbf{q}_\mathrm{IC}$}

%
% rez lattice vectors
\newcommand{\ra}{$\mathbf{a}^*$}
\newcommand{\rb}{$\mathbf{b}^*$}
\newcommand{\rc}{$\mathbf{c}^*$}

%
% Fermi energy
\newcommand{\EF}{$\epsilon_\mathrm{F}$ }
%
% x-ray / X-ray
\newcommand{\xray}{X-ray }
%
% chemical formular
\newcommand{\ttas}{1T-$\mathrm{TaS_2}$}
%
% short cuts
\newcommand{\eg}{e.\,g}
\newcommand{\ie}{i.\,e}

\title{Pressure dependence of the charge density wave in 
1T-TaS$_\mathbf{2}$ and its relation to superconductivity}

\author{T.\,Ritschel}
\affiliation{Leibniz Institute for Solid State and 
Materials Research IFW Dresden, Helmholtzstrasse 20, 01069 Dresden, Germany}
\affiliation{Institute for Solid State Physics, Dresden Technical University,
TU-Dresden, 01062 Dresden, Germany} 
\author{J.\,Trinckauf}
\affiliation{Leibniz Institute for Solid State and 
Materials Research IFW Dresden, Helmholtzstrasse 20, 01069 Dresden, Germany}

\author{G.\,Garbarino}
%\author{A.\,Bossak}
\author{M.\,Hanfland}
\affiliation{ESRF Grenoble, France}

\author{M.\,v.\,Zimmermann}
\affiliation{Hamburger Synchrotronstrahlungslabor HASYLAB at Deutsches
Elektronensynchrotron DESY, Notkestr. 85, 22603 Hamburg, Germany}

\author{H.\,Berger}
\affiliation{Ecole polytechnique Federale de Lausanne, Switzerland}

\author{B.\,B\"uchner}
\affiliation{Leibniz Institute for Solid State and Materials Research IFW
Dresden, Helmholtzstrasse 20, 01069 Dresden, Germany} 
\affiliation{Institute for Solid State Physics, Dresden Technical University,
TU-Dresden, 01062 Dresden, Germany} 
\author{J.\,Geck}
\email[Author to whom correspondence should be addressed: ]{j.geck@ifw-dresden.de}
\affiliation{Leibniz Institute for Solid State and Materials Research
IFW Dresden, Helmholtzstrasse 20, 01069 Dresden, Germany}

\date{\today}

\begin{abstract}
  We present a state-of-the-art x-ray diffraction study of the charge density
  wave order in \ttas{} as a function of temperature and pressure.
  %
%  The wave vector, the amplitude and the coherence length of the charge density wave are determined in different regions of the phase diagram 
  %
  Our results prove that the charge density wave, which we characterize in terms
  of wave vector, amplitude and the coherence length, indeed exists in the
  superconducting region of the phase diagram. 
  %
  %The pressure dependence of
  %the wave vector, the amplitude and the coherence length of the charge density wave, .  
  %
  The data further imply that the ordered charge density wave structure as
  a whole becomes superconducting at low temperatures,  \ie , superconductivity
  and charge density wave coexist on a macroscopic scale in real space. This
  result is fundamentally different from a previously proposed separation of
  superconducting and insulating regions in real space and, instead, provides
  evidence that the superconducting and the charge density wave gap exist in
  separate regions of reciprocal space.
\end{abstract}

%
%\pacs{75.85.+t, 77.80.-e, 64.70.Rh, 61.05.C-}
% 75.85.+t : Magnetoelectric effects, multiferroics
% 77.80.-e : Ferroelectricity and antiferroelectricity
% 64.70.Rh : Commensurate-incommensurate transitions
% 61.05.C- : X-ray diffraction and scattering

\maketitle

%%%%%%%%%%%%%%%%%%
%  introduction  %
%%%%%%%%%%%%%%%%%%

It is an intriguing fact that in many complex materials superconductivity --a
state of matter where charge can move through a lattice without any resistance--
often exists in close proximity to what appears to be exactly the opposite: the
static spatial ordering of charge. The very recently reported charge density
wave (CDW) instability, which competes with superconductivity (SC) in the
cuprates is only one example\,\cite{Ghiringhelli2012,Chang2012}. Similar issues are
also discussed for the newly discovered iron pnictide
superconductors\,\cite{Zhang2009a}, heavy fermion systems\,\cite{Gegenwart2008}
and the dichalcogenides\,\cite{Gabovich2002,Du2000}. The relation between electronic order
and SC therefore receives considerable attention. But despite all research
efforts, the question if or under which circumstances electronic order competes,
coexists or supports superconductivity remains mostly controversial.

Here we investigate the relation of CDW and SC in the layered binary material
\ttas{}. This compound provides a well suited model system, as its underlying lattice structure is simple and the CDW order at
ambient pressure has already been characterized in
detail\,\cite{Spijkerman1997,Wilson1975a}. Interestingly, in \ttas{} a cascade of different CDW
transitions occurrs: with decreasing temperature, an
incommensurate (IC) CDW develops first at about 550\,K, which, upon further
cooling, changes into a nearly commensurate (NC) CDW at $\approx 350$\,K. This
phase then finally transforms into a commensurate (C) CDW below 190\,K, which is
commonly described as a Mott-phase due to electron-electron
interactions\,\cite{Fazekas1980,Fazekas1979}. Indeed, very recent time-resolved
measurements revealed ultra-fast charge dynamics in the C-CDW
phase\,\cite{Perfetti2006,Hellmann2010a,Eichberger2010,Petersen2011}, 
lending support to the notion that
electron-electron interactions are important and implying that the C-CDW is
beyond conventional electron-phonon physics.

\begin{figure}[b!]
  \begin{center}
    \includegraphics{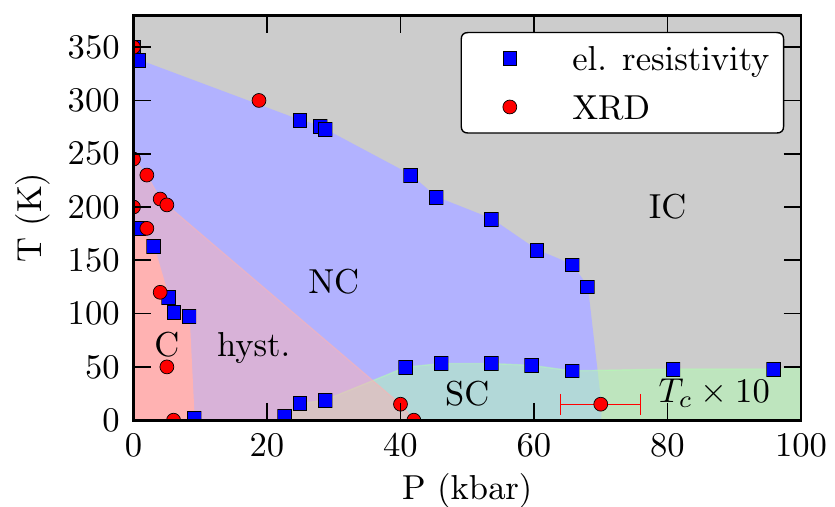}
  \end{center}
  \caption{(color online) The pT-phase diagram of \ttas{}. Blue square markers represent data
    from electrical resistivity measurements\,\cite{Sipos2008} and red circle
    markers the transition temperatures found by XRD in this study (C:
    commensurate, NC: nearly commensurate, IC: incommensurate, SC:
    superconducting).  The C phase is characterized by a large hysteresis
    illustrated by the transparent  reddish region.
  }
  \label{fig1}
\end{figure}

The occurrence of all these electronic phase transitions and in particular the
appearance of a Mott-phase is already very interesting. But not long ago it was
found that pristine \ttas{} also becomes superconducting below 5\,K at pressures
above 40\,kBar\,\cite{Sipos2008}.  The rich electronic phase diagram  of
\ttas{} as a function of pressure (p) and temperature (T) is shown in \myref{fig1}, where we reproduce  previous results along with data from this
study.

\begin{figure*}[t!]
  \centering
  \includegraphics{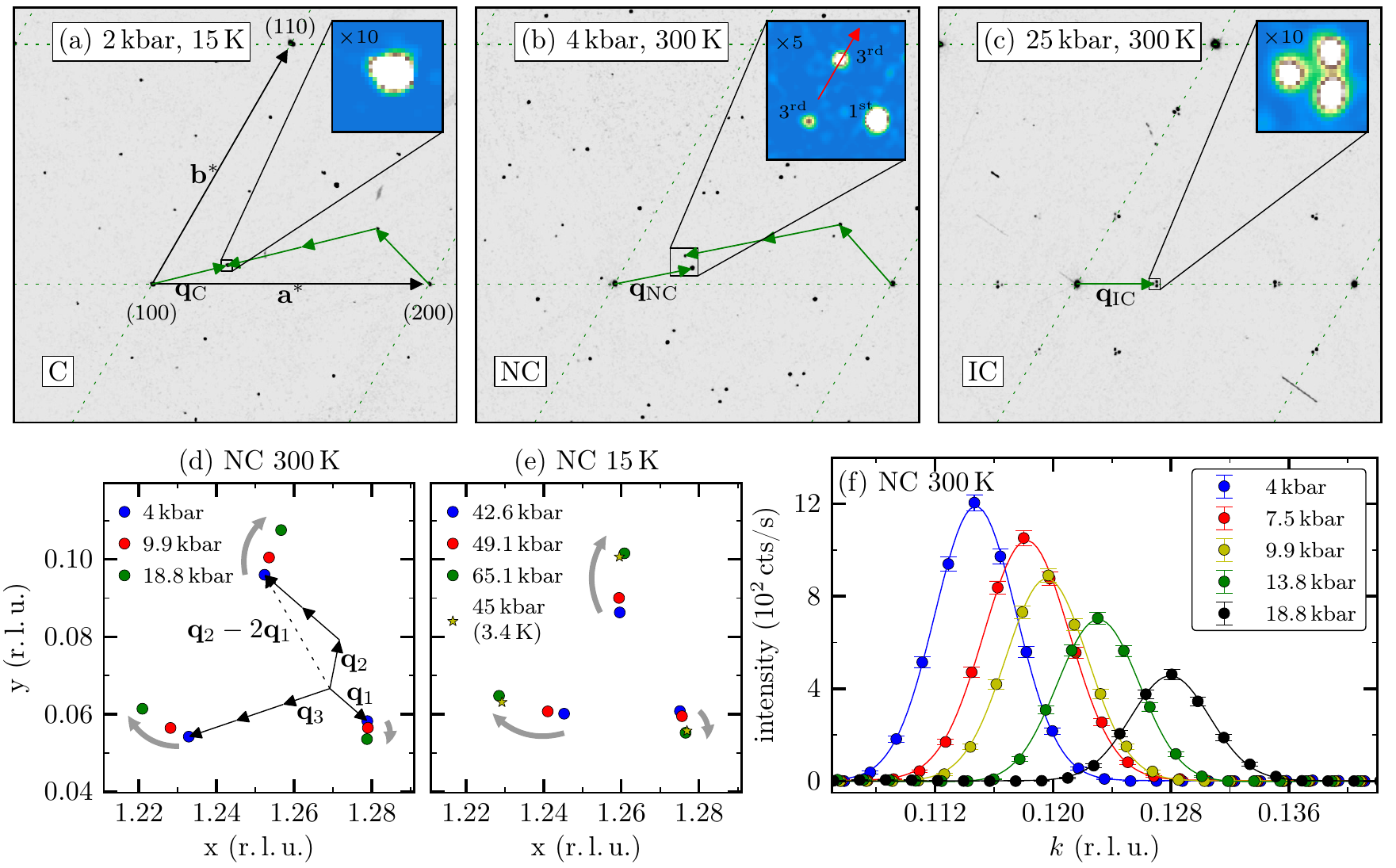}
  \caption{(color online) Reciprocal space maps of the XRD intensity for the C-phase (a), the
    NC-phase (b) and the high-pressure IC-phase (c).  In (a) the Bragg reflections are indicated by the Miller indices $(hkl)$ and the reciprocal lattice vectors  $\mathbf{a}^*$, $\mathbf{b}^*$ of the hexagonal plane are shown by black arrows. A magnified region in
    reciprocal space is displayed in the insets, where the threefold splitting
    of the satellite reflections in the NC-phase and the high pressure IC-phase
    can be clearly observed .  (d) and (e) show the satellite peak positions in
    the NC  phase as a function of p at 300\,K and 15\,K, respectively. In (d)
    $\mathbf{q}_1 = \mathbf{q}_\mathrm{NC} - \mathbf{q}_\mathrm{C}$ and
    rotating $\mathbf{q}_1$ by $120^\circ$ and $240^\circ$ yields $\mathbf{q}_2$
    and $\mathbf{q}_3$, respectively. At both 300\,K and 15\,K, the modulation
    wave vector clearly shifts towards the IC-position with increasing p. (e)
    also includes data taken in the SC region of the phase diagram at 45\,kbar
    and 3.4\,K.  (f)  $k$-scans  through the 3$^\mathrm{rd}$ order satellite
    peak (along red arrow in inset of (b))  versus p at T=300\,K, illustrating
    the clear p-dependence of the peak position and intensity. The solid lines
    represent fitted pseudo-Voigt profiles.
  }
  \label{fig2}
\end{figure*}

Based on detailed resistivity measurements,  Sipos \textit{et.~al.} deduced
a microscopic scenario for SC in \ttas, according to which SC develops in metallic region that separate insulating C-CDW domains and grow with increasing
pressure\,\cite{Sipos2008}. In other words, the pressure induced SC and its
coexistence with CDW order was explained in terms of a microscopic phase
separation in real space.
% into metallic and insulating domains, where SC develops only within
%the metallic regions.  
It is clear, however, that macroscopic  measurements
cannot provide information about the microscopic spatial structure of the CDW.
Since this information is essential in order to understand the coexistence of SC
and CDW in \ttas{}, we scrutinized the CDW order by means of x-ray diffraction
(XRD) experiments as a function of  p and T.

%%%%%%%%%%%%%%%%
%  Experiment  %
%%%%%%%%%%%%%%%%

The high-quality single crystals used for the present XRD study were grown by
the the iodine vapor transport method as described in
Ref.~\onlinecite{Zwick1998}.  The majority of the present XRD measurements were
conducted at the beamline ID09 of the ESRF. Oriented samples of about 80\,$\mu$m
diameter were loaded in a membrane driven diamond anvil pressure cell filled
with helium as the pressure transmitting medium. For the low-temperature
measurements the pressure cell was then installed in a continuous He-flow
cryostat and exposed to a $10\times10$\,$\mu$m$^2$ beam with a photon energy of
30\,keV.   A \textsf{MAR555} flat panel detector was used to collect the
diffraction data in large regions of reciprocal space. At each pressure, we
collected a dataset of 120 images over a sample rotation of 60$^\circ$ with
0.5$^\circ$ scan width per image.  
We increased the pressure up to 150\,kbar and 80\,kbar at constant temperatures
of 300\,K and 15\,K, respectively, and monitored the pressure \textit{in
situ} using the ruby fluorescence.
During the low-temperature measurements we also cooled the sample to 3.5\,K at
every pressure point above 40\,kbar, in order to reach the superconducting
phase.  In addition to these measurements, the C-NC transition at lower
pressures was investigated at beamline BW5 of DESY. Here we used a clamp-type
pressure cell\,\cite{Zimmermann2008} and performed measurements as a function of
temperature at constant pressure.

%%%%%%%%%%%%%
%  Results  %
%%%%%%%%%%%%%

The reflections observed in XRD enable to determine the spatial arrangement of
the lattice sites in a solid. In a CDW material one generally observes Bragg
reflections, which are related to the underlying average structure. The CDW
induces additional modulations of that structure and since the period of the CDW
in real space is larger than that of the underlying lattice, 
additional reflections appear around the Bragg peaks. These are referred to as
superlattice or satellite reflections.
The position, intensity and width of the satellite reflections provides direct
information about the spatial structure, the amplitude and the correlation
length of the CDW. 

The  XRD intensity was recorded as a function of the scattering vector
$\mathbf{Q}$, which is commonly given in terms of the Miller indexes $(hkl)$:
$\mathbf{Q}=h\,\mathbf{a}^*+k\,\mathbf{b}^*+l\,\mathbf{c}^*$ with
$\mathbf{a}^*$, $\mathbf{b}^*$ and $\mathbf{c}^*$ the reciprocal lattice vectors
of the unmodulated structure (cf.\,\myref{fig2}(a)). Since the satellite
reflections in \ttas{} occur at different none-zero $l$-values\,\cite{Spijkerman1997}, we integrated the scattered intensity along the
$l$-direction, resulting in diffraction pattern that correspond to projections
of the x-ray intensity within a slice of thickness $\Delta l = 2/3$  onto the
$hk0$-plane in reciprocal space. Typical XRD datasets obtained in this way are
presented in \myref{fig2}, where the additional satellite reflections around
every Bragg peak can be clearly observed. 

The diffraction pattern taken at 300\,K and close to ambient pressure is shown
in \myref{fig2}(b). Under these conditions the NC-CDW exists that is
characterized by a wave vector \ncqs{}, which deviates slightly from the
commensurate wave vector \cqs. As illustrated in \myref{fig2}\,(b), the slight
incommensurability of \ncqs{} results in two third order satellites close to
the first order peak. The observation of strong higher order satellite
reflections verifies that the NC-phase is characterized by a domain like
structure with sharp
boundaries\,\cite{Spijkerman1997}.
As can be seen in \myref{fig2}(a), the incommensurability and the resulting
splitting of the satellite peaks vanishes in the C-CDW Mott-phase, which is
reached when the sample is cooled down while keeping p close to ambient pressure
(vertical path close to p=0\,kbar in \myref{fig1}). This phase is characterized
by a commensurate wave vector \cqs{} and is stabilized by electron-electron
interactions\,\cite{Fazekas1980,Tanda1984}.

Keeping T constant at RT and increasing p, the IC-phase is reached in agreement
with earlier reports (horizontal path close to T=300\,K in \myref{fig1}).
However, as shown in the inset of \myref{fig2}(c), the pressure-induced IC-phase
differs from the one at ambient pressure in that it shows an additional
splitting of the satellite reflections within the $hk$-plane. This observation
is in accordance with a previous study\,\cite{Ravy2012}, but the reason for this
splitting is still unclear and subject of ongoing investigations.  
%In the IC
%phase only first order satellite reflections are visible, which indicates
%a rather smooth CDW modulation. 

The effect of increasing p at constant T on the wave vector \ncqs{} is presented
in \myref{fig2}(d) and (e).  We determined the peak positions by fitting
2D-Gaussian profiles to the measured diffraction pattern, including up to 50
first order and 25 third order satellite reflections. This enabled us to
determine the peak positions with high accuracy. As can be observed in
\myref{fig2}(d) and (e), the position of the satellite reflections clearly
changes upon increasing p, which corresponds  to \ncqs{} moving towards \icqs. 
For geometrical reasons the shift in position is more pronounced for the
third order satellite reflections, as can be clearly seen in \myref{fig2}(d)
and (f).  The shift in position together with the behavior of the peak width and
intensity is illustrated in  \myref{fig2}(f), which shows  scans along the
reciprocal $k$-direction. Not only the peak shifts according to the change of
\ncqs{}. Also the intensity of the reflection is strongly suppressed, revealing
a pronounced reduction of the CDW amplitude. The analysis of the peak profiles,
however, does not show any significant broadening of the peaks, \ie, no change
in the coherence length of the CDW is observed.

\begin{figure}[t!]
  \begin{center}
    \includegraphics{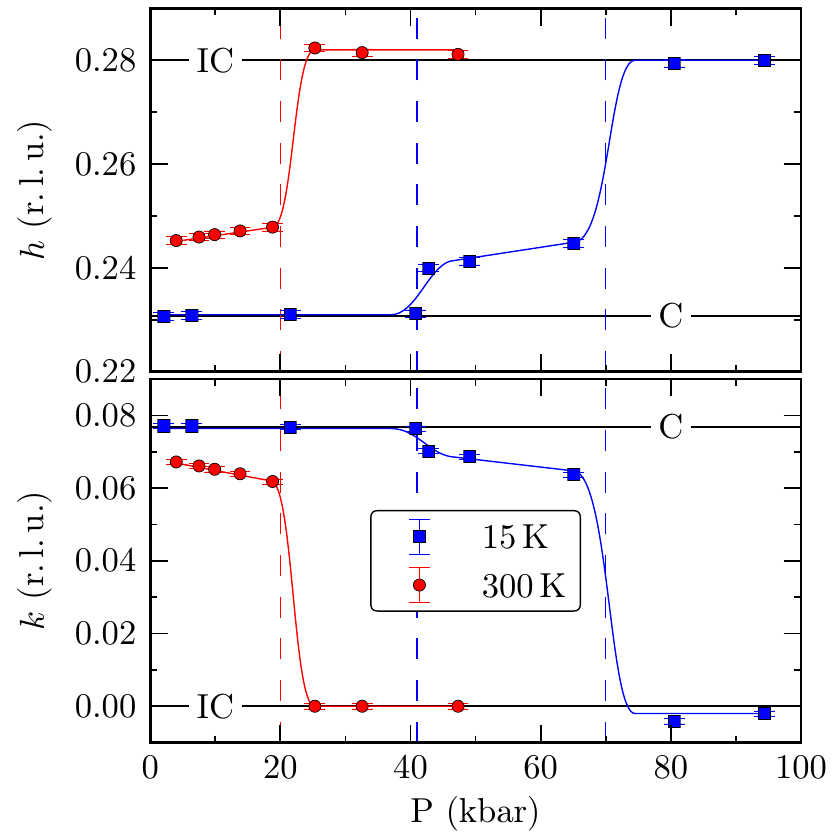}
  \end{center}
  \caption{Inplane components of the modulation wave vector as a function of
    pressure.  Red Closed circle markers and blue square markers represent the
    room and low temperature measurement, respectively. At room temperature the
    NC-IC-transition occurs at $P\approx 20$\,kbar (red dashed line). For low
    temperatures we observed the C-NC-transition and the NC-IC-transition at
    about 40\,kbar and 70\,kbar, respectively (blue dashed lines). Within the
    NC-phase the $\mathbf{q}$-vector shifts towards \icqs. Solid lines are
    guides to the eye.} \label{fig3}
\end{figure}

The in-plane components of the modulation wave vector determined by the fitting
procedure are also summarized quantitatively for the RT and 15\,K measurement in
\myref{fig3}.  Starting with the low-temperature data set, the C-NC transition
is observed close to 40\,kbar with increasing p. Within the NC-phase, the \ncqs{}
moves clearly towards \icqs{} without reaching it completely. Then at about
70\,kbar a sudden jump of the modulation vector to \icqs{} signals a first order
transition into the IC-phase (the threefold splitting is neglected and only the
midpoint is shown). A corresponding behavior of the NC-IC transition is also
observed at room temperature. These data agree very well with the pT-phase
diagram deduced from resistivity measurements as illustrated in \myref{fig1}. 

It is remarkable that the CDW at 15\,K remains commensurate up to 40\,kbar, when
the pressure is increased at constant T=15\,K, because the C-phase is suppressed
already at 6\,kbar for temperature sweeps at constant p. This was shown by
resistivity measurements\,\cite{Tani1977,Sipos2008} and also verified by our
XRD. Both experiments also showed a very large difference for the transition
temperatures, depending on whether the sample is cooled or heated at constant p.
These observations imply that the C-phase is metastable in a large
pressure-temperature region (cf.\,\myref{fig1}).

In order to search for possible changes of the CDW order in the SC-phase, we
also cooled the sample from 15\,K down to $\approx 3.5$\,K at every pressure
point within the NC-phase.  However, no significant change of the CDW order
could be detected upon entering the SC-region of the phase diagram (cf. star marker in
\myref{fig2}(e)). Note, that upon cooling to the lowest T, the pressure
decreased as well. Yet no significant changes of the CDW order were observed,
which we attribute to hysteresis effects. 

%%%%%%%%%%%%%%%%
%  discussion  %
%%%%%%%%%%%%%%%%

We now turn to the discussion of the presented XRD results and their
implications for the coexistence of SC and CDW in \ttas. As already mentioned in
the introduction, the p-induced
superconductivity in this compound was recently explained in terms of a phase
separation scenario, which is based on the microscopic structure of the
NC-phase\,\cite{Sipos2008}. The NC-CDW at ambient p is characterized by
hexagonal shaped C-CDW domains separated by domain
boundaries, which are also called
discommensurations\,\cite{Spijkerman1997,Thomson1994,Ohta1992,%
Wu1989,McMillan1975,McMillan1976,Nakanishi1977}.
The latter are commonly regarded as charged and metallic regions in-between C-CDW
domains\,\cite{Sipos2008}.
However, it is important to realize that this is not a domain structure in the
usual sense, because the C-CDW domains have a well-defined shape and size and,
importantly, their spatial arrangement is periodically ordered. As a result, the C-CDW domains
and the domain boundaries together form a regular kagome lattice with a large
coherence length, yielding sharp satellite reflections in reciprocal space. 

\begin{figure}[t!]
  \centering
  \includegraphics{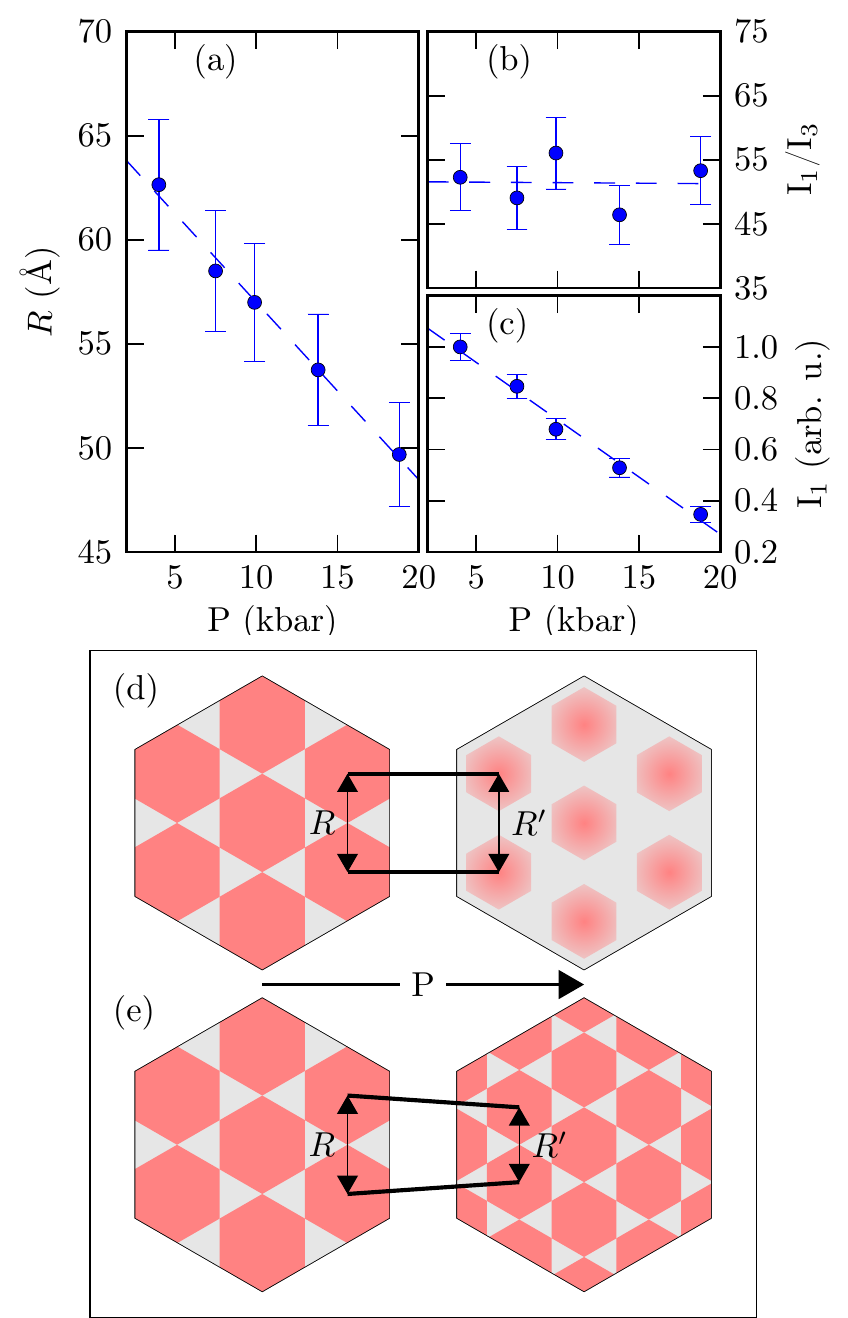}
  \caption{(a): Calculated C-domain distance as a function of pressure using
    Eq.\,\ref{eqn:domainsize}.  (b): Intensity ratio between the first and third
    order satellite reflection. (c): Overall intensity of the first order
    satellite reflection. Dotted lines are guides to the eye. (d) and (e)
    illustrate two possible scenarios of pressure induced C-domain shrinking in
    real space. The reddish hexagons refer to C-CDW domains and the gray areas
    represent the discommensurations. (d): the distance between C-CDW domains
    remains constant and the domain boundaries smear out. In (e) the C-domain
    distance shrinks and the domain boundaries remain sharp.}
  \label{fig4}
\end{figure}

As it was shown previously\,\cite{Nakanishi1977}, the average distance $R$ of
neighboring C-CDW domains (see \myref{fig4}) is directly related to the
incommensurability of the CDW via
\begin{equation} 
  R =
  \frac{8\pi}{3\sqrt{13}\cdot|\mathbf{q}-\mathbf{q}_{\mathrm{C}}|},
  \label{eqn:domainsize} 
\end{equation}
where $R$ is given in lattice units,
$\mathbf{q}$ is the measured modulation vector and \cqs\/ is the modulation
vector of the commensurate phase. In addition to this, the sharpness of the
C-domain boundaries determines the intensity ratio
between first and higher order satellite reflections. 

According to the scenario proposed in Ref.~\onlinecite{Sipos2008}, the
insulating C-CDW domains shrink with increasing pressure and, hence, the metallic
domain boundaries widen and become interconnected as sketched in
\myref{fig4}(d). At a certain pressure, the superconductivity can eventually
occur at low temperature within the connected metallic regions.
In this scenario, $R$ will remain essentially constant with increasing pressure.
Furthermore, the widening of the domain walls corresponds to smooth boundaries
between neighboring C-CDW domains, which will result in a substantial change of the
intensity ratio between the first and third order satellite reflection
$\mathrm{I_1/I_3}$.  As can be seen in \myref{fig4}(a) and (b)  these two
characteristic changes are not observed. Instead, we find a clear reduction of $R$ and
a constant $\mathrm{I_1/I_3}$ ratio within the errors of the experiment. Our
data therefore does not agree
with the scenario illustrated in \myref{fig4}(d).

%According to our results, $R$ shrinks with increasing p and, 
%which means that the C-CDW domains shrink as well.  at the same time, 
The constant $\mathrm{I_1/I_3}$ shows that the boundaries between neighboring C-CDW domains remain
sharp, while $R$ and with it the size of the C-CDW domains shrink with
increasing p. Our results hence imply that the spatial structure of the CDW
changes as illustrated in \myref{fig4}(e). Further, the data in \myref{fig4}(c)
shows that
the intensity of the satellite reflections decreases by a factor of 3, \ie., the
overall amplitude of the lattice modulation decreases by  $\approx 1/\sqrt{3}$
with pressure.

The shrinking of the C-CDW domains and the reduction of the CDW-amplitude observed
by XRD agrees with the conclusions reported previously in
Ref.\,\onlinecite{Sipos2008}. The important new result here is that the domain
boundaries in the NC-phase do not form large interconnected metallic regions.
We also do not observe a dissociation of the C-CDW domains, which would result
in a strong broadening and, eventually, the disappearance of the
NC-superlattice reflections. Instead, the sharp XRD peaks in the NC-phase prove
that the metallic regions and the C-CDW domains in this phase always form
a long-ranged ordered and periodic structure.

The pressure-induced formation of large metallic regions therefore seems not to
be crucial for the SC in \ttas. Rather, the behavior illustrated in
\myref{fig4}(e) requires that the ordered  structure as a whole becomes
superconducting. In other words, not only the metallic regions support SC, but
the whole NC-structure illustrated in \myref{fig4}(e) forms a coherent
macroscopic superconducting state.
The same conclusion was also reached for the SC-phase that is induced in \ttas{}
by Fe-substition. From a completely different view point, namely that of
angle-resolved photoemission spectroscopy, Ang {\it et al.} also found that the
NC-CDW and SC must coexist in real space\,\cite{Ang2012}.
Our results for the p-induced SC together with the study of Fe-induced SC by Ang
{\it et al.} provide solid experimental evidence for SC occurring in the NC-CDW
structure as a whole; a situation which is fundamentally different from the
previously proposed phase separation in real space.
The essentially p-independent superconducting transition temperature
$\mathrm{T_c}$, together with the p-induced changes of the NC-CDW observed here,
further implies that NC-CDW and SC are not competing. 
%
%This indicates that the electronic states, which are involved in the SC are only little affected by the pressure induced changes of the NC-CDW. 
We therefore argue that instead of a phase separation in real space there is
a phase separation in k-space: NC-CDW gap and SC gap occur in separate regions
of the Fermi surface. 

While, according to the pT-phasediagram, the same should also be true for the
IC-CDW, the Mott C-CDW clearly competes with SC. Most likely because it
completely gaps the Fermi surface and hence leaves no states for the
superconducting condensate. Further dedicated studies of the electronic
structure as a function of pressure are however necessary to verify these
conjectures. In particular, notwithstanding the obvious competition between the
Mott C-CDW and SC, it remains to be clarified whether or not electron-electron
interactions are relevant for the SC of \ttas. We believe that the
superconducting CDW in \ttas{} can serve as a viable model, which will help to
understand also other complex materials, sharing the same pathology.

This work was financially supported by the German Research Foundation under
grant DFG-GRK1621.  J.T and J.G gratefully acknowledge the financial support by
the German Research Foundation through the Emmy Noether program (grant
GE\,1647/2-1). We thank A.~Bosak for fruitful discussions.

%\bibliography{mybib}

%%%%%%%%%%%%%%%%%%%%%%%%%%
%  ./HXRD_1T-TaS2_revised.bbl  %
%%%%%%%%%%%%%%%%%%%%%%%%%%


\end{document}